\documentstyle[12pt]{article}

\newcommand{\be}{\begin{equation}}
\newcommand{\ee}{\end{equation}}
\newcommand{\bea}{\begin{eqnarray}}
\newcommand{\eea}{\end{eqnarray}}

\newcommand{\I} {{\cal H}}

\newcommand{\M} {{\tilde M}}
\newcommand{\V} {{\cal V}}

\begin{document}

\begin{center}
\begin{large}
{\bf  Bad News \\}  
{\bf on the \\}  
{\bf Brane \\}
\end{large}  
\end{center}
\vspace*{0.50cm}
\begin{center}
{\sl by\\}
\vspace*{1.00cm}
{\bf A.J.M. Medved\\}
\vspace*{1.00cm}
{\sl
Department of Physics and Theoretical Physics Institute\\
University of Alberta\\
Edmonton, Canada T6G-2J1\\
{[e-mail: amedved@phys.ualberta.ca]}}\\
\end{center}
\bigskip\noindent
\begin{center}
\begin{large}
{\bf
ABSTRACT
}
\end{large}
\end{center}
\vspace*{0.50cm}
\par
\noindent

\par
There has been substantial  interest in  obtaining a
   quantum-gravitational  description of de Sitter space. However, 
any such attempts have encountered formidable obstacles, and
  new philosophical directions may be in order.
One possibility, although  somewhat  speculative,
would be to view the physical
universe as a timelike hypersurface evolving in a higher-dimensional
bulk spacetime; that is, the renowned brane-world scenario.
In this paper, we extend some  recent studies
along this line, and consider a non-critical 3-brane
moving in the  background
of  an anti-de Sitter  Reissner-Nordstrom-like 
black hole. Interestingly, even an arbitrarily small
electrostatic charge in the bulk can induce a 
singularity-free ``bounce'' universe on the brane, whereas a vanishing
charge typically implies  a singular cosmology.
 However, under closer  examination,
from a holographic (dS/CFT)  
perspective, we  demonstrate that the charge-induced
bounce cosmologies  are  not physically viable.
This implies the necessity for censoring against
charge in a  bulk black hole.

\newpage

\section{Introduction}

Recent empirical evidence  of an  accelerating universe  
\cite{bops}   is most easily explained
in terms  of a  positive vacuum energy density or cosmological
constant.
  This observation has triggered substantial
interest in de Sitter (dS) space, which  happens to 
be the maximally symmetric solution of Einstein gravity
when this constant takes on a positive, non-vanishing value.
\par
In particular, one hopes that a quantum   theory
of dS  gravity can eventually be realized.
Work along this line  has been inspired
by  significant achievements  in the realm of anti-dS space
(i.e., maximally symmetric
 Einstein gravity with a negative cosmological constant).
For instance, higher-dimensional spacetimes in string theory 
(our best candidate for
a fundamental theory of quantum gravity)  can often be
decomposed into
 anti-dS space times a simple compact manifold.
(For a review on string theory, see \cite{PEET}.)
Furthermore, there is a well-known correspondence
between  any anti-dS spacetime and a conformal field theory
(CFT) that ``lives'' at  spatial infinity \cite{mal,gub,wit}. 
Significantly,  this duality is a direct manifestation
of the holographic principle \cite{tho,sus,bouxx}, which is expected
to  be a fundamental constituent in
the ultimate theory of quantum gravity.
\par
Unfortunately, the  remarkable achievements of anti-dS
space have not translated well into a dS setting.
For example,  attempts at compactifying string theory
into a stable vacuum with a positive cosmological constant
have been plagued by pathologies; including
singularities, wrong-sign kinetic terms and
non-compact ``compactification'' manifolds. (See \cite{dds}
and references therein for further discussion.)
Meanwhile, from a holographic perspective,
there has, actually, been significant progress in establishing
a dS analogue\footnote{A
 dS/CFT holographic duality was first rigorously
formulated in \cite{str}. For both earlier and later
studies on this topic, one can consult 
the bibliographies of \cite{med2,medyyy,NEW}.
More recent studies of interest include 
\cite{BOUXX,CRAP,MCIX,DEH,BOUX,NOO,CCCM,DAN}.}
 to the anti-dS/CFT correspondence.
However, the status of this conjectured dS/CFT duality
still remains a very open question \cite{dls}.
\par
Progress along the desired lines has been stymied
by  various ``detrimental'' features of dS space;
including the lack of an asymptotic spatial infinity,
the absence of an objective observer,  the existence
of an observer-dependent event horizon, the lack of a global
timelike killing vector, the eternally thermal background, {\it etcetera}.
(For further discussion on the properties
of dS space, see \cite{SVV,lmmx}.) 
However, it is the  finite entropy of dS space \cite{gh2}
that appears to be the main  impediment towards a quantum gravity
realization \cite{banks,bou,bfxx,witxx}.
Significantly, this finite bound on the  entropy 
implies that any dS theory of quantum gravity 
should be described by a finite-dimensional Hilbert space.
Alas, this deduction  is in  conflict with the infinite-dimensional
Hilbert space associated with string theory,
any of its  related
 conceptualizations  (such as
matrix theory \cite{matrix})
and, from a holographic perspective, any CFT.
 That is to say, the  finite entropy of dS space  appears
to sabotage  any prospective  quantum-gravity  framework
 at a very fundamental level.
\par
With regard to the above conundrum,  it is of further
interest
 that the finite entropy of dS space serves
as an upper bound on the observable entropy of
a large class of spacetimes with a positive cosmological
constant; including any  spacetime with an asymptotically
dS  future \cite{bou}.
\par
 It should be quite  clear that any
 assimilation of quantum  physics and
dS gravity  will  have to overcome some very formidable obstacles.
Although speculative, one possible road to salvation may yet 
arise out of
the realm of brane-world scenarios \cite{ruby}.
For  sake of definiteness, let us now outline
a  particular form of   
brane-world  model, as inspired by Randall and Sundrum \cite{rs},
 which will be adhered to throughout  the paper.
In the scenario of interest,  our physical universe is 
 regarded as a  3+1-dimensional hypersurface,
or 3-brane, immersed in a 4+1-dimensional anti-dS
bulk spacetime (with any additional dimensions having
been compactified to string-length scale).
We also follow the usual convention of 
confining all standard-model particles to the brane
except for the graviton.
In compliance with the observed cosmological constant \cite{bops},
we further stipulate a non-critical
brane with positive curvature.\footnote{Although the graviton  
propagates freely into the bulk,
it has been shown that gravity will typically remain localized even
when the brane is curved \cite{wow, psnd}.}
Furthermore, following Kraus \cite{kra}, we take the viewpoint
of a brane that flows through an otherwise static, black hole
bulk.
\par
Given the above framework,
there are a couple of points of interest that should be kept
in mind.  Firstly, the brane dynamics will effectively
describe the cosmological evolution of the universe.
That is to say, from the perspective of an observer,
the  motion of the brane will appear as either
a cosmological contraction or expansion.
Secondly, the (effective) cosmological constant on the brane
can be expressed in terms of a pair of bulk parameters;
namely, the anti-dS curvature radius and the brane tension.
 In this sense, the cosmological
constant can be regarded as an input parameter
\cite{banks}, rather than a  variable  quantity
of the lower-dimensional gravity theory.
\par
One might well query  as to how  this brane-world interpretation
 could possibly  address 
the difficulties associated with  a quantum formulation
of dS gravity. The answer lies in the notion of
a physical universe that  can now be   holographically viewed
in  an  anti-dS framework \cite{hmsxxx,ver}. As  discussed earlier, the
quantum-gravitational aspects of anti-dS spacetimes are
much better understood than their dS counterparts. 
Hypothetically speaking,  one could  obtain a  string-theoretical
description of the $5$-dimensional bulk spacetime 
and then apply the anti-dS/CFT duality  to
translate this theory onto the brane \cite{hver}. Although, admittedly,
such  a prospect remains somewhat speculative at the present time.
\par
In  the current treatment, we  investigate the 
above brane-world scenario
for the intriguing case of a charged (anti-dS) black hole in the bulk.
Before discussing the
 content of this paper, let us take note of some recent works  
of relevance. First of all, Petkou and Siopsis \cite{pet} considered
the cosmological and holographic implications of a brane world evolving
in the background of an anti-dS Schwarzschild black hole. 
This study was subsequently generalized by the  present 
author \cite{NEW} to include ``topological'' 
anti-dS black holes \cite{bir}. That is, Schwarzschild-like black
holes
(static with a constant-curvature horizon), but
having an  arbitrary horizon topology (flat, hyperbolic or spherical).
\par 
More recently, Mukherji and Peloso \cite{MUPE}
have considered topological anti-dS black holes with a finite
electrostatic charge (i.e.,  Reissner-Nordstrom-like).\footnote{For
some earlier
studies and discussion on brane cosmology with a bulk charge, see  
\cite{xxx,bmw,AA1,yyy,AA2,AA3,AA4,AA5}.}
These authors focused their analysis on  a critical brane
(i.e., the effective cosmological constant  is
fine tuned to vanish); although they briefly
touched upon the qualitative features of a non-critical
brane scenario as well. In the program to follow,
we elaborate on this last  case of a  non-critical (positively curved)
brane evolving in a charged black hole background.
Along with the switch in emphasis (from critical to non-critical),
  the current study substantially  deviates
from  \cite{MUPE} in the following manner:
we come here not to praise charged black hole bulks 
but, rather, to bury them.
\par
The remainder of the paper is organized as follows.
In Section 2,
we introduce the relevant formalism; including the
bulk solutions of interest (topological anti-dS black holes
with charge)
 and the equation of motion
for the brane. Notably, this equation mimics
the Friedmann equation for radiative
matter, along with an additional exotic form of  matter.
This  latter contribution, which only arises 
when the bulk solution has a finite charge, can be identified
as  stiff matter with a  negative-energy density \cite{bmw}.
\par
 In Section 3, we consider the cosmological implications
for a certain class of solutions of the priorly
mentioned Friedmann equation. Although explicit analytical
expressions are out of our reach, it is still possible 
to describe the cosmologies of interest. Here, we
 appeal to  asymptotic regimes for which
 the relevant  solutions  have  already been formulated
\cite{NEW,MUPE}. In this process, we observe that a non-vanishing  charge in
the bulk (even an arbitrarily small one) generally induces a FRW ``bounce''
 universe; that is, asymptotically dS and devoid of 
singularities.\footnote{For general discussion on Friedmann-Robertson-Walker
(FRW)  bounce
cosmologies, see \cite{lmmx,kklt}. For a  recent, topical
study, from both a brane-world and string-theoretical perspective,
 see \cite{BBB}.}
Conversely, for the class of solutions under investigation,
a vanishing charge implies a singular spacetime; either
a ``big bang'' or a ``big crunch''.
\par
In Section 4, we consider the holographic implications
of the charged-induced  bounce cosmologies.  More precisely, we  calculate
their generalized $c$-functions \cite{lmmx,kklt}  (as  prescribed 
for renormalization group flow in an asymptotically 
dS spacetime \cite{STR2,bdm,HAL})
and then determine how these functions evolve in time.
On the basis of this analysis, we deduce that
any such  bounce cosmology, although free of singularities,
is  of an unphysical nature.  We attribute this
failure  to the exotic, negative-energy matter  
that is induced by the bulk charge. Here, we note 
that   similar failures have been observed 
when negative-energy matter is introduced
into an otherwise purely dS spacetime \cite{MCI,medyyy}.
\par
Finally, Section 5 contains a summary and  considers
future prospects.

\section{Brane-World Scenario}

The formal interest of this paper is a 
Randall-Sundrum \cite{rs} type of  brane-world scenario; more specifically, 
 a 3+1-dimensional brane (of positive tension)
moving in a 4+1-dimensional bulk spacetime that is otherwise
static. Furthermore, we will assume that the bulk geometry
is described by an  anti-dS black hole with an electrostatic charge
and a constant-curvature horizon. 
The relevant
black hole solutions can be regarded as ``Reissner-Nordstrom-like'',
but with an arbitrary horizon topology.
\par
The   solutions in the anti-dS
 bulk can thus be expressed as follows \cite{RNBH}:
\be
ds^2_{5}=-f(r)dt^2+{1\over f(r)}dr^2+r^2d\Omega^2_{k,3},
\label{2}
\ee
where:
\be
f(r)= {r^2\over L^2}+k-{\omega M\over r^{2}}
+{3\omega^2 Q^2\over 16 r^4},
\label{3}
\ee
\be
\omega={16\pi G_5\over 3\V}.
\label{4}
\ee
Here,   $d\Omega^2_{k,3}$  denotes the line element of a 3-dimensional
constant-curvature (Euclidean) hypersurface,
 $\V$ is the dimensionless volume
of this
hypersurface, $L$ is the curvature radius of
the anti-dS bulk (i.e., $\Lambda_{5}=-3/L^2$ is the
bulk cosmological constant) and $G_5$ is
the 5-dimensional Newtonian constant. There are also
three constants of integration in this solution:
 $k$, $M$ and $Q^2$.
Without loss of generality, $k$ can  be set equal  to +1, 0 or -1;
describing a horizon geometry that is respectively
spherical (i.e., the  anti-dS Reissner-Nordstrom case),
flat or hyperbolic.
Meanwhile, $M$ and $Q^2$ respectively measure the conserved
mass and  the
 electrostatic charge (squared) of the black hole; with these being
regarded as strictly non-negative quantities.\footnote{It should be noted
that the $k=-1$ hyperbolic solution supports a negative value
for the mass \cite{bir}. However,  this 
controversial scenario will be disregarded, on the grounds that a negative-mass
 black hole  probably induces a non-unitary
boundary theory  \cite{caixxx}.}
\par
Note that the existence of a pair of  positive, real  horizons or
non-extremal black hole solution
(as we assume to be the case in this analysis)
depends on, in general, a sufficiently large value of $L^2$
and, for $k=+1$,  a sufficiently small value of $Q^2$
(in fact, $Q^2 < 4 M^2/3$).\footnote{One can verify these claims
 by solving for the roots of Eq.(\ref{3}).
These roots, of course, locate the positions of the horizons.}
To be definite, we will assume that 
$M$  is ``large'' (with $L^2$ always taking on a
large enough value to compensate)
and $Q^2$ is ``small''. More precisely, we  henceforth limit
the following dimensionless parameters:
$\M > 1$ (where $\M\sim M$ is defined below)
and $\epsilon^2\equiv  3 Q^2 /4 M^2 < 1$ (and, typically, $\epsilon^2<<1$).
That is, the  black hole of interest is appropriate
for  a semi-classical regime and has a   small
(but non-vanishing) electrostatic charge. 
\par
As  demonstrated elsewhere (for instance, \cite{kra}),  if 
one considers an evolving
 brane in an otherwise static bulk,  the  dynamics of the brane
will mimic that of a
FRW universe \cite{frw}.
Here, we will  skip the gory details (see \cite{NEW} for
a recent presentation) and simply quote the  results.
\par
As it so happens, the induced metric on the brane
can be expressed  in the following
 FRW form:
\be
ds^2_{4}=-d\tau^2+r^2(\tau)d\Omega^2_{k,3},
\label{9}
\ee
where $\tau$  measures the physical time
from the point of view of a brane observer  and
$r$ is the time-dependent cosmological scale factor.
Moreover, the corresponding Friedmann equation
is  found to be as follows:
\be
H^2= {\Lambda_{4}\over 3}-{k\over r^2}+{\omega M\over r^4}
-{\omega^2 M^2 \epsilon^2\over 4 r^6},
\label{16}
\ee
where $H\equiv {\dot r}/r$ is the usual Hubble ``constant''
(with a dot denoting differentiation with respect to
$\tau$) and $\Lambda_{4}$ is the effective cosmological
constant on the brane. In achieving this form,
we have invoked the following defining relation:
\be
\Lambda_{4}\equiv 3\left[\left({\sigma\over 3}\right)^2
-{1\over L^2}\right],
\label{17}
\ee
where $\sigma$ represents the tension of the brane (see
\cite{NEW} for a precise definition).
\par
To touch base with the status  of
our own universe \cite{bops}, we will assume
that $\Lambda_{4}$ takes on a relatively small,  positive  value.
From a brane-world standpoint, this translates into
a non-critical, positively curved brane universe.\footnote{Traditionally,
the Randall-Sundrum brane-world scenario \cite{rs} assumes a
fine tuning of the brane tension
so that $\Lambda_{4}$ vanishes; that is, a critical brane.
However, there is no {\it a priori} rationale
for this choice of brane tension, and we note
some prior studies 
\cite{wang,noxxx,pet,pad1,youmxxx,medxxx,pad2,NEW} that
have considered the dynamics of a non-critical brane.}
Note that the  horizon geometry of the black hole
(i.e., the choice of $k$) directly determines
the topology of the spatial slicing in the brane universe. 
\par 
The  above form (\ref{16}) can, in fact, be identified  with
the 4-dimensional Friedmann equation  
for radiative matter (since $\rho_{rad}\sim r^{-4}$,
where $\rho$ denotes energy density)  along with
an  exotic matter contribution by  virtue of
the $r^{-6}$ term. This latter contribution can
be identified with stiff matter  (which is defined by $p_{sti}=\rho_{sti}$,
where $p$ denotes pressure) \cite{bmw}.  What makes
this holographic stiff matter all the more intriguing
is its  negative prefactor in Eq.(\ref{16}); thus implying
that  $\rho_{sti}<0$.
 Although  negative-energy  matter is typically
frowned upon, as it directly violates
 any number of positive-energy conditions \cite{WALD},
we note that such  conditions have their foundation in
technical convenience rather than fundamental
arguments \cite{VIS}. That is to say, it is still too
early in the game to base  the physical status
of a theory on this criteria alone. Hence, we will keep
an open mind (for the time being) and  go on to consider
the implications of including this exotic matter.
\par
Although we have clearly established a brane-world pedigree
for the relevant cosmological equation (\ref{16}),
one can bypass the brane description altogether  and  view this 
picture from a strictly
4-dimensional sense. That is to say, one may
regard  Eq.(\ref{16}) as the Friedmann equation for a toy cosmological model;
namely, a model universe that contains radiative matter, exotic
stiff matter and a positive cosmological constant.
From this perspective, one can view the current work as another  chapter in
 ``How {\it not} to construct an asymptotically de Sitter universe''
\cite{medyyy}.

\section{FRW  Cosmology on the Brane}

Ideally, we would like to obtain an analytical solution
to Eq.(\ref{16}) for reasonably  generic circumstances.
Although this does not seem to be possible, it
is  still instructive to consider the asymptotic limits
(i.e.,  very large or very small $r$) for
which (approximate) analytic expressions are indeed obtainable.
As it turns out, the asymptotic behavior of the 
scale factor is, by itself, sufficient for our current purposes. 
\par
First, let us consider the solutions for the special circumstance
of vanishing charge
(i.e., $Q^2$ or $\epsilon^2=0$).
For this special case, we turn  to a recent work \cite{NEW}
(also see \cite{pet})
and quote the  relevant expressions for all three
choices of $k$:
\\
(i) $k=+1$:
\be
r^2={1\over 2\I^2}\left(1+\sqrt{\M-1}\sinh\left[\pm 2\I(\tau-\tau_o)\right]
\right).
\label{31}
\ee
(ii) $k=0$: 
\be
r^2= {\sqrt{\M}\over 2 \I^2}\sinh\left[\pm 2\I\tau\right].
\label{32}
\ee
(iii) $k=-1$:
\be
r^2={1\over 2\I^2}\left(\sqrt{\M-1}\sinh\left[\pm 2\I(\tau-\tau_o)\right]
-1\right).
\label{33}
\ee
Here, we have incorporated the following definitions: 
${\cal H}^2 \equiv \Lambda_{4}/ 3$, 
$\M\equiv 4\omega M{\cal H}^2$, and $\tau_o$
is a  constant of integration that can always be chosen so
that $r^2$ vanishes at $\tau=0$ (for sake of convenience).
Keep in mind that we have assumed ${\cal H}^2$ to be positive
and  $\M>1$. (Solutions for other choices of $\M$ are
given in \cite{NEW}.)
\par
Some discussion on the above solutions is in order.
Each solution  describes a FRW universe that either
begins at a ``big bang'' 
or ends at a ``big crunch'', depending on the choice of sign
in the hyperbolic argument (positive or negative, respectively).
These two scenarios are 
effectively equivalent, by way of time-reversal symmetry,
so it  is sufficient to illustrate  just the big bang universe.
In this case, the  brane world  ``comes into existence''
at $\tau=0$, where  $r=0$ and the curvature can
be shown to diverge.  As time proceeds, the scale factor
monotonically increases and ultimately  describes
an asymptotically dS universe ($r\sim e^{\I\tau}$)
in the far future ($\tau\rightarrow\infty$). 
It is interesting to note that the asymptotic solution
is essentially  ignorant of the details
of the bulk geometry.
\par
What makes the above solutions particularly significant
is their general validity  for very large values of $r$
(or, equivalently, large values of $|\tau|$). In this
regime, we can neglect the effects of charge,
given that the charge term in Eq.(\ref{16})
falls off rapidly ($\sim r^{-6}$) as the scale
factor increases. Hence, as long as  there
is no singularity in the spacetime (which is always, as shown below,
to be true  for non-vanishing charge), the scale  factor
must asymptotically approach  the above forms in either the distant
past or far future.\footnote{Note that the sign in the hyperbolic functions
should always be suitably chosen to ensure $r^2>0$.}
\par
Next, let us consider the solutions for another special case; namely,
$\Lambda_{4}=0$ or a critical brane. For these solutions, we
turn to a recent analysis by Mukherji and Peloso \cite{MUPE}. 
These authors obtained the following results:
\\
(i) $k=+1$:
\be
r^2={\omega M\over 2}\left(1-\sqrt{1-\epsilon^2}\cos\left[2\eta\right]
\right).
\label{31.5}
\ee
(ii) $k=0$: 
\be
r^2=  {\omega M\over 4}\left(\epsilon^2+4\eta^2\right).
\label{32.5}
\ee
(iii) $k=-1$:
\be
r^2={\omega M\over 2}\left(\sqrt{1+\epsilon^2}\cosh\left[ 2\eta\right]
-1\right).
\label{33.5}
\ee
Note that the above have been expressed in terms of
conformal time, $\eta$, which is defined
according to $d\tau=r d\eta$. Also note that $\epsilon^2 >0$
has been assumed.
\par
A very interesting feature of these critical solutions
is that they all have a non-vanishing minimum value
for the scale factor (which occurs in the above forms,
by way of arbitrary convention,  precisely at $\eta=0$).
That is to say, as long as the charge is non-vanishing,
a singularity will certainly be avoided. To put it another
way, these critical solutions all ``bounce''
(at least)  at the surface of vanishing  time.
 In fact, the $k=+1$ solution
bounces, by virtue of
its periodicity,
 an infinite number of times.
\par
As for the prior case  of  vanishing charge,
the special nature  of a critical brane
belies the generality of the above solutions.
On inspecting the Friedmann equation (\ref{16}), we see that, for
small enough values of the scale factor, the
$\Lambda_4$ term will make a negligible contribution.
More to the point, $\Lambda_4$ is a finite-valued constant, 
 whereas all
the other terms in Eq.(\ref{16}) increase rapidly as the
scale factor decreases. Hence, any relevant solution   
must asymptotically approach the above forms 
(\ref{31.5}-\ref{33.5}) as
$\eta$ goes to zero. Therefore, given a bulk black
hole with a non-vanishing charge, there will always
be a non-vanishing minimum  for the scale factor on the brane.
In fact, this singularity-free status
 holds up for
arbitrarily small values of $\epsilon^2 \sim Q^2$. 
\par
In spite of the conspicuous absence of a generic
analytical solution, we are now in a position
to say something about the brane cosmologies
of interest (i.e., those with $\Lambda_4>0$, $1>\epsilon^2 >0$, and $\M>1$).
It is clear that, for at least the cases of a flat ($k=0$) and open ($k=-1$)
brane universe, the scale factor will have an asymptotically
dS form  in the distant past, contract to a non-vanishing  minimum
at $\tau=0$, and then expand towards an asymptotically  
dS future. Such a cosmological scenario
is often said to describe a FRW bounce universe. The closed-universe
($k=+1$) case is, however, not so clear. It is not evident
whether the scale factor, after its initial bounce, will oscillate between
a maximal and minimal value (as suggested by Eq.(\ref{31.5})),
or manage  to ``escape''  towards an asymptotically dS 
 future (as implied by Eq.(\ref{31})).
The ultimate outcome depends, in all likelihood, on
the relative values of the model-dependent parameters; in particular,
 $M$ and $\Lambda_4$. One can imagine that, for a sufficiently small
value of $M$, the vacuum energy will dominate and
an exponential expansion will ensue. Conversely,  if
$M$ is above  some threshold value, one might expect
that the radiative matter wins out and the brane universe
eternally oscillates. It would be interesting
to verify the existence (or lack thereof) of such a threshold
value; however, a numerical analysis is almost certainly
required.
\par
It is quite interesting 
that  even an arbitrarily small amount
of negative-energy matter 
 can have such  a dramatic effect on the
 underlying cosmology. 
When the bulk charge is precisely vanishing (i.e., no  exotic matter
on the brane),
 a  bounce cosmology will {\it never} be possible;  assuming that
 $\M>1$ has also been enforced.\footnote{In fact, for vanishing charge,
a  closed  brane universe can only bounce 
if $\M<1$ \cite{pet} and an open/flat universe will only
bounce if $\M=0$ \cite{NEW}.}
That is, under these conditions,
  a singular surface (either a big bang or big crunch)
is an inevitable feature of  the spacetime manifold.
Conversely, after introducing
arbitrarily small  amounts  of negative energy  onto the  brane,
 we   find that
 the  universe is regular throughout  and (in most cases) 
has both an asymptotically
dS past and  future.  
\par
On some intuitive level, the above outcome is not  particularly
surprising.  Regardless of how small  (a non-vanishing) $\epsilon^2$
is chosen to be,  it is clear  from
the Friedmann equation (\ref{16}) that the negative-energy
term will dominate at sufficiently small values
of the scale factor (thanks to the $r^{-6}$ factor
in this stiff-matter term).
Hence, as the brane universe contracts towards $r=0$,
the negative-energy matter  must, at some point, take over and create a 
significant repulsive force; thus
 preventing the otherwise inevitable collapse of the spacetime.
\par
In spite of the ambiguous status \cite{VIS} of the various
positive-energy conditions \cite{WALD}, one might be troubled
by a universe that is dominated by negative-energy matter
at some time in its evolution.
Moreover,  the resilient nature of these bounce cosmologies
 is most puzzling, insofar as 
any asymptotically dS spacetime
should have   an upper bound on its observable entropy 
(\cite{bou} and see Section 1).
That is to say, there is
clearly  no upper limit  on the size of  $M$ and, hence,
presumably no   bound on the amount of entropy
that can be ``pumped'' onto the brane.
Therefore, at least naively, one would expect that the dS entropy bound
can easily be violated.
\par
Fortunately, these 
unsettling matters  can still be resolved without
 bringing the positive-energy conditions or  entropic upper 
bound directly
 into play.
For this purpose, we next call on our old friend, holography.

\section{Holographic Interpretation}

Essential to the following analysis is the concept of
holographically induced renormalization group (RG) flows.
Let us, therefore, give a brief account of this phenomenon.
\par
Firstly, we  consider RG flows as they apply to 
the well-known anti-dS/CFT  
correspondence \cite{mal,gub,wit}.
In this anti-dS  context, it has been established
that the monotonic evolution of
a relevant bulk parameter induces a ``flow'' in
the renormalization scale of the dual boundary theory,
and {\it vice versa}. (Keep in mind that this
renormalization scale determines the ultraviolet cutoff or,
equivalently, the lattice spacing of the boundary theory.)
This framework follows, in large part,
from the so-called ultraviolet/infrared correspondence
\cite{UVIR1}. That is, high (low) energies in the boundary
theory translate into large (small) distances 
in the anti-dS bulk.
\par
Significant to any RG flow is the existence of
a generalized $c$-function. Such a function follows,
by way of analogy, with  the $c$-function of
a 2-dimensional CFT \cite{ZAM}. For   anti-dS holography in particular,
 any such  $c$-function is expected to
exhibit various monotonicity properties  that are
reflective of the ultraviolet/infrared duality.
For further  pertinent discussion, see (for instance)
\cite{AG,FGPW,SAH,BVV}. 
\par
Next, we will extend the above discussion to  
a dS  holographic framework. 
It is first worth recalling that, in analogy
to  anti-dS holography,
 a  dS spacetime has been conjectured
to  have a dually related boundary theory \cite{str}.\footnote{For
the pathway to other literature on this subject matter, see
Section 1.}
Such a  boundary theory, if it does exist, would
presumably be a Euclidean CFT that lives  at one or both of
past and future infinity.  On the other hand, 
the dS/CFT correspondence is not without its detractors
(most notably, \cite{dls}\footnote{And for the inevitable
counter-argument, see \cite{KLEMM}.}). However, in spite
of the unclear status of this duality,
there is little doubt that  many aspects of anti-dS  holography do indeed
translate over to a dS setting; including  RG flows.
\par
Let us now (finally!) focus  on RG flows in a dS context.
As observed by Strominger \cite{STR2} (also see \cite{bdm}), 
time evolution in a purely dS  spacetime  will generate
conformal-symmetry transformations on its asymptotic spacelike boundaries.
However, this conformal symmetry  will be broken
if  the spacetime is ``demoted'' to being only  asymptotically dS.
 On the basis of these observations, the author  went
on to argue that  time evolution (in an asymptotically dS universe) 
will  naturally induce a RG flow between a pair of conformal
fixed points.
These fixed points occur in the asymptotic past and future when
the symmetries of pure dS space ultimately re-emerge.
Moreover, in analogy to anti-dS holography,
 Strominger proposed an associated $c$-function
of the form:
\be
c\sim \left|\left({{\dot r}(\tau)\over r(\tau)}\right)^{-(n-1)}\right|,
\label{weee}
\ee
where
$r$ is the FRW scale factor, a dot denotes differentiation
with respect to cosmological time ($\tau$), and
 $n+1$ is the spacetime dimensionality. 
\par
Significantly to its status as a legitimate  $c$-function,
the above form does indeed  exhibit  appropriate
monotonicity properties  \cite{STR2,bdm}.
In particular, $c$ flows to the ultraviolet (infrared) - that is, 
 increases (decreases) - for an expanding (contracting) universe.
\par
One issue of concern is that Eq.(\ref{weee}) can only
be applied to  a  FRW
 spacetime with flat spatial slices (i.e., $k=0$).
Nonetheless,  the above  form of $c$
has since been generalized so that all choices of
spatial slicing can indeed be utilized 
\cite{lmmx,kklt}.\footnote{In particular,
Leblond {\it et al} \cite{lmmx} proposed the generalized form
 and  verified the appropriate monotonic behavior,
while Kristjansson and Thorlacious \cite{kklt}
provided a geometrical interpretation by identifying
$c$ with the area of an apparent horizon.}
This generalization can be expressed as follows:
\be
c\sim \left({r^2(\tau)\over k+
{\dot r}^2(\tau)} \right)^{(n-1/2)},
\label{blahh}
\ee
where $k$ is the usual FRW topological parameter. 
 It is easy to
see that this form reduces to the prior one when 
$k=0$.
\par
To reclarify, $c$ is expected to increase if
 the universe is expanding  and decrease
if the universe is contracting. An interesting technical
issue   is the  application of this ``$c$-theorem''
to a bounce universe, as such a cosmology 
obviously has both an expanding and a contracting phase.
In this sense, one is forced to abandon the
conventional wisdom that a $c$-function should evolve
in a strictly monotonic fashion. (For further
discussion on this caveat, see \cite{lmmx}.)
\par
By this time, the reader may  be wondering as to the point of this
minor dissertation on holographic RG flows and the like.
Well, as it so happens, RG flows have recently been  utilized 
(specifically by McInnes \cite{MCI}; also  see \cite{medyyy})
to expose certain types of  bounce cosmologies 
as being physically unacceptable. 
Interestingly, these charlatan cosmologies also
contain  negative-energy matter.  However, unlike
the current study, no other form of matter was
included (besides a cosmological constant).
In spite of this fundamental difference,  we will  proceed to show 
that the same
debilitating argument applies  here as well.   
\par
For sake of calculational simplicity, let us consider the case
of an open brane universe or flat-horizon black hole (i.e., $k=0$).
We will restrict considerations to the expanding phase
($\tau$ or $\eta >0$), as time-reversal symmetry
enables  one  to choose either expansion or contraction
without loss of  generality.
First of all, let us focus on the large $r$ (or late  time) solution,
as presented in Eq.(\ref{32}). Substituting this result into
the above form of the $c$-function (\ref{blahh}),
we find  that:
\be
c\sim {\sinh^2\left[2\I(\tau-\tau_o)\right]\over \cosh^2\left[
2\I(\tau-\tau_o)\right]},
\label{111}
\ee
where we have set $n=3$ and neglected any irrelevant
constant factors.
\par
Our primary objective is to ascertain how $c$ responds to temporal variations.
Given the generic positivity of the $c$-function, it
is sufficient (and considerably simpler) to consider variations in
$\ln[c]$.  Again neglecting constant (positive) factors,
we have:      
\be
{\partial \ln[c]\over \partial \tau}= 
{1\over \sinh\left[4\I(\tau-\tau_o)\right]} > 0.
\label{222}
\ee
Evidently, this expression satisfies the expectations of
the $c$-theorem; namely, $c$ should  monotonically
increase with time during an expanding phase of the universe.
\par
Next, let us consider  the small $r$ (or small $|\tau|$)
solution, as documented in Eq.(\ref{32.5}). 
Before proceeding, we should first re-express
the  $c$-function (\ref{blahh}) 
in a form that is directly compatible with
the conformal-time variable, $\eta$. That is:
\be
c\sim \left({r^4(\eta)\over kr^2(\eta)+
 \left[r^{\prime}(\eta)\right]^2} \right)^{(n-1/2)},
\label{blahhh}
\ee
where a prime indicates differentiation with respect to
conformal time.
\par
Substituting Eq.(\ref{32.5}) into the above, we obtain: 
\be
c\sim {1\over \eta^2}\left(\epsilon^2 +4\eta^2\right)^3.
\label{333}
\ee
Again  varying the logarithm (and  recalling
that $\eta>0$ has, without loss of generality, been assumed),
we find that:
\be
{\partial \ln[c]\over \partial \eta}= {12\eta\over \epsilon^2 +4\eta^2}
-{1\over \eta}.
\label{444}
\ee
\par
It is difficult to make sense of this result,
until we remember that this asymptotic solution (\ref{32.5}) 
is only 
 strictly valid  for  vanishingly small values of $\eta$.
Hence, it is most appropriate to consider the limit of
 $\eta\rightarrow 0_+$,
in which case:
\be
{\partial \ln[c]\over \partial \eta} \rightarrow -{1\over 0_+} <0.
\label{555}
\ee
This time around, we observe a direct violation of the
relevant $c$-theorem.
That is, for small values of $r$,
 $c$ monotonically decreases with time 
 during an expanding phase of
the universe. This outcome implies that time
flows in the ``wrong''  direction when the brane universe is small;
assuming that
Strominger's interpretation  
-  time evolution is dual with a RG flow to the ultraviolet \cite{STR2} - 
 should be taken
literally. 
\par
Before elaborating on this curious behavior,  we point out
that the above qualitative features will naturally
persist for both the closed ($k=+1$) and open ($k=-1$)
brane-universe scenarios.
This  realization follows from  an
inspection of Eq.(\ref{16}),  from  which it is clear
that the topological ($k$) term will not play
a significant role in either asymptotic regime.
\par
Generally speaking, a violation in the $c$-theorem, although
definitely a ``red-flag'',  should
not  necessarily  be regarded as an unphysical outcome.
In support of this claim, we point out 
that  dS holography is not particularly  well understood
(not in comparison to its anti-dS analogue),
and it remains uncertain as to how literally
its predictions should be interpreted.
Moreover, the validity of the $c$-theorem depends
on the assumption of the ``weak energy condition'' 
\cite{WALD,STR2,bdm,lmmx},
which will certainly be violated  whenever negative-energy
matter is dominant. (For our brane model, this domination occurs 
 when the  universe is small.)
With this caveat in mind, one could well have anticipated
the observed failure of the $c$-theorem at small
values of $r$. To put it another way, the  preferred direction
of time flow is not particularly clear until the second
law of thermodynamics can somehow be invoked.
\par
In spite of this attempt at spin-doctoring, there 
remains  a  pressing concern that cannot, quite so  easily, be 
argued away:  not only does time apparently
flow in the  wrong direction, but it actually reverses
direction at some point in the evolution of the universe.
That is, the $c$-function  flows to the infrared
when the universe is small but ends  up 
flowing to the ultraviolet at some later time.
Such a ``phase transition'' in the $c$-function\footnote{This behavior
seems
oddly reminiscent of the Hawking-Page anti-dS phase
transition \cite{HP,wit}. It  may be of interest 
to investigate this possible analogy; however,
such a line is impeded by our lack of a solution for the non-asymptotic 
 universe.}
does not appear to be compatible with {\it any}
interpretation of holographic RG flows; taken literally
or otherwise. That is to say, we can perhaps tolerate a RG flow
that reverses direction at a bounce, but {\it not} one that
reverses direction while the universe is in the process
of expanding (or contracting). 
\par
 In view of
the preceding discussion, we are forced to conclude
that these  bounce cosmologies with exotic stiff matter 
are physically unacceptable on holographic grounds.  
Moreover, from a brane-world  perspective, this failure implies
that a bulk black hole should be prohibited
from acquiring an electrostatic charge.

\section{Conclusion}

In summary, we have been studying  the  cosmological and
holographic  implications 
of a certain class of  brane-world scenarios.  To elaborate,
we have focused on a non-critical (positively curved) 3-brane moving
in an  anti-dS background that is described by a 
Reissner-Nordstrom-like  black hole.
It should be noted that our analysis
 incorporated the  topological variants of
the ``standard'' (anti-dS)  Reissner-Nordstrom solution;
thus allowing for spherical, flat and hyperbolic
 horizon geometries.  
\par
We began the analysis by introducing the relevant bulk solutions
and the equation of motion for the brane.
It is significant that the form of the latter
essentially  mimics the  Friedmann equation
for radiative matter.
Along with the radiative contribution (which is proportional
to the mass of the bulk black hole) and an  effective
cosmological constant,
this induced Friedmann  equation   
contains an additional exotic form of matter.
This exotic contribution, which is holographically induced
by the bulk electrostatic charge, can
 be identified as stiff matter with a negative energy density. 
Also of interest, the horizon geometry of the black hole
fixes  the spatial topology 
of  the brane universe.
\par
In the next section  of the analysis, we
specifically considered  bulk black holes
with  a relatively large mass and a relatively small
(but non-vanishing) charge. The interest in this regime follows
from   prior studies on uncharged black holes in the bulk \cite{pet,NEW}. 
These earlier treatments have indicated 
that a cosmological singularity is inevitable when the black hole mass
exceeds a certain (topology-dependent) value.
That is,  under these conditions, the  brane universe
must either begin with a big bang or end in a big crunch.
\par
Although   the induced
Friedmann equation is  not generally  solvable,  we were still able
to proceed by  focusing on  a pair of asymptotic regimes.
In fact,  the solutions for very large and very small values
of the scale factor are known (\cite{NEW} and \cite{MUPE}, respectively),
and these  could be used  to deduce the  gross features
of the relevant cosmologies (up to some
ambiguity in the case of a closed brane universe).
In precisely this manner, we have demonstrated that even an arbitrarily
small charge (in the bulk) is sufficient to avoid a singularity in
the brane universe. Rather than collapsing to a singularity
(as is inevitable in the uncharged  scenario),
the brane universe either bounces towards an asymptotically
dS future  (as is always the case for a flat or open universe)
or eternally oscillates between a maximal and minimal
size. 
\par
In the final phase  of  our  study,
we investigated the charge-induced bounce cosmologies
from a holographic perspective. More to the point,
we calculated the generalized $c$-functions
(as   prescribed for dS-holographic RG  flows
 \cite{STR2,bdm,lmmx,kklt})
in the asymptotic regimes and then  tested
these functions against the recommended $c$-theorem.
Notably, this methodology was inspired by
prior treatments \cite{MCI,medyyy} that revealed a violation
of the $c$-theorem  for a special class
of bounce cosmologies. Significantly, these
cosmologies also contain  negative-energy matter.
\par
Ultimately, in applying this litmus test,
we have  demonstrated that the charge-induced
bounce cosmologies are of an unphysical nature. More
precisely, when the universe is very small
(and the negative-energy matter dominates),
the associated $c$-function  decreases as the
universe expands; an outcome  which is in direct violation of
the $c$-theorem.  Although one could still
argue on how literally the $c$-theorem
should be interpreted (given that the direction
of time is perhaps ambiguous in a universe with built-in time-reversal
symmetry),  we  pointed out an even  more disturbing  consequence
of this calculation. 
Namely,  the  RG flow must reverse its direction,  at some point,
while the universe is  expanding. (This follows from
the behavior of the  $c$-function in the
other asymptotic regime, where it clearly increases
as the universe expands.) 
Such a reversal  is decidedly in conflict with
the philosophical premise  of  holographic RG flows.
\par
In view of the above arguments, we conclude that
the charge-induced bounce universes are, indeed, unphysical
cosmologies.  Moreover, assuming the feasibility
of brane-world scenarios in general,
we suggest the necessity  for censoring
against charge in a bulk black hole.
\par
Of course, the last declaration may be somewhat premature,  inasmuch
as  we still have no direct knowledge of the non-asymptotic evolution.
 It would be
interesting to identify the ``reversal point''
and   better understand whatever  mechanism is deviously at work.
However, it would appear that such an investigation
would require a numerical analysis.  As an alternative
to this unsettling proposition,
one might consider the scenario of a 2-brane moving
in a 4-dimensional charged black hole bulk.
Although unphysical, this simpler model
may have an analytical solution with similar 
features to that of the priorly studied case.
\par
One might also extend the prior treatment by considering
other FRW cosmologies that contain
negative-energy matter (as well as a positive cosmological constant).  
Clearly, a pattern
is arising:  the presence of such exotic matter
consistently leads to violations in the holographic $c$-theorem.
On the other hand, the  models so-far studied  have
a common feature; namely, the negative-energy
matter dominates when the universe is very small.
It may be of interest to construct an analytically
solvable model in which this is not the case.
Given the ambiguous status of the various
positive-energy conditions \cite{VIS}, it  would certainly
be useful to ascertain their feasibility 
 via such  independent means.
Naturally, we defer the above prospects to a future
(cosmological or conformal) time.

\section{Acknowledgments}
\par
The author  would like to thank  V.P.  Frolov  for helpful
conversations.



\end{document}